\documentclass[aip,superscriptaddress,reprint, nofootinbib]{revtex4-1}%
\usepackage{epsfig,pslatex,latexsym,times,amssymb,amsmath,graphicx}
\usepackage{bm, float, lipsum}
\usepackage{amsmath}
\usepackage{amsfonts}
\usepackage{amssymb}
\usepackage{mathrsfs, bbold}
\usepackage{color}
\usepackage{graphicx}%

\usepackage[normalem]{ulem}

\providecommand{\U}[1]{\protect\rule{.1in}{.1in}}
\begin{document}
\author{Nicholas J. Harmon}
\email{harmon.nicholas@gmail.com} 
\affiliation{Department of Physics, University of Evansville, Evansville, Indiana
47722, USA}
\author{James P. Ashton}
\affiliation{Department of Engineering Science and Mechanics, Pennsylvania State University, University Park, PA, 16802, USA}
\author{Patrick M. Lenahan}
\affiliation{Department of Engineering Science and Mechanics, Pennsylvania State University, University Park, PA, 16802, USA}
\author{Michael E. Flatt\'e}
\affiliation{Department of Physics and Astronomy and Optical Science and Technology Center, University of Iowa, Iowa City, Iowa
52242, USA}\date{\today}
\title{Near-Zero-Field Spin-Dependent Recombination Current and Electrically Detected Magnetic Resonance from the Si/SiO$_2$ interface}
\begin{abstract}
Dielectric interfaces critical for metal-oxide-semiconductor (MOS) electronic devices, such as the Si/SiO$_2$ MOS field effect transistor (MOSFET), possess trap states that can be visualized with electrically-detected spin resonance techniques, however the interpretation of such measurements has been hampered by the lack of a general theory of the phenomena. 
This article presents such a theory for two electrical spin-resonance techniques, 
electrically detected magnetic resonance (EDMR) and the recently observed near-zero field magnetoresistance  (NZFMR), by generalizing Shockley Read Hall trap-assisted recombination current calculations via  stochastic Liouville equations. Spin mixing at this dielectric interface occurs via the hyperfine interaction, which we show can be treated either  quantum mechanically or semiclassically, yielding distinctive differences in the current across the interface.  By analyzing the bias dependence of NZFMR and EDMR, we find that the recombination in a Si/SiO$_2$ MOSFET is well understood within a semiclassical approach.
\end{abstract}
\maketitle	

\section{Introduction} 

Magnetic resonance experiments have given access to the microscopic details of defects inside various materials, including semiconductors and insulators, through electron spin resonance (ESR) techniques, primarily the technique referred to as electron paramagnetic resonance (EPR).\cite{Wertz1970} Increased sensitivity is provided by the closely associated ESR methods called optically and electrically detected magnetic resonance (ODMR and EDMR).\cite{Chen2003, Boehme2005} These latter methods take advantage of spin-selection rules so while they are more limited in their use (since they require transitions between spin pairs) they are also detectable through sensitive optical and electrical measurements. For all these methods a radio frequency or microwave frequency field is required to induce the spin transitions. 
Magnetic resonance techniques have been particularly useful in discovering and determining properties of deep-level paramagnetic defects in semiconductors and insulators that are widely used in contemporary and prospective integrated circuits.\cite{Lenahan1998, Watkins1999} For instance, the P$_b$ defect, a dangling bond, appears at the interface of Si and SiO$_2$ in Si/SiO$_2$ MOSFETs (where it is identified as P$_{b0}$);\cite{Nishi1971, Nishi1972} these centers capture charge carriers, shifting the threshold voltage, and reduce effective transistor channel mobilities.\cite{Lenahan1982, Lenahan1984, Kim1988, Vranch1988, Miki1988, Awazu1993} If these defects are present in significant numbers, which occurs when irradiated or stressed by other means, their presence substantially limits the performance of the transistors.\cite{Ashton2019, Fleetwood2009}


The electronic states associated with P$_b$ and the variant P$_{b0}$ lie near the middle of the band gap, and thus contribute substantially to recombination current, allowing them to be accessed efficiently using EDMR.  Although the defects were first observed with conventional electron paramagnetic resonance (EPR)  measurements,\cite{Nishi1972} the sensitivity of conventional EPR, about ten billion total defects in the sample under study,\cite{Eaton2010} is not sensitive enough for studies of these centers at the dielectric interfaces embedded within technologically meaningful small devices.
Since the sensitivity of EDMR measurements is at least ten million times higher than that of conventional EPR, such measurements allow resonant investigations of these fundamental materials interfaces in fully processed technologically meaningful devices so long the interface locations are reachable by a.c. magnetic fields. 
EDMR measurements have demonstrated that   P$_b$ centers  play important roles in determining the response of MOSFETs to multiple technologically important device   stressing phenomena including the injection of hot carriers into the oxide\cite{Krick1991},  the effects of negative gate bias at elevated temperature,\cite{Campbell2007} and radiation.\cite{Ashton2019} 
\begin{figure}[ptbh]
 \begin{centering}
        \includegraphics[scale = 0.6,trim = 0 0 0 0, angle = -0,clip]{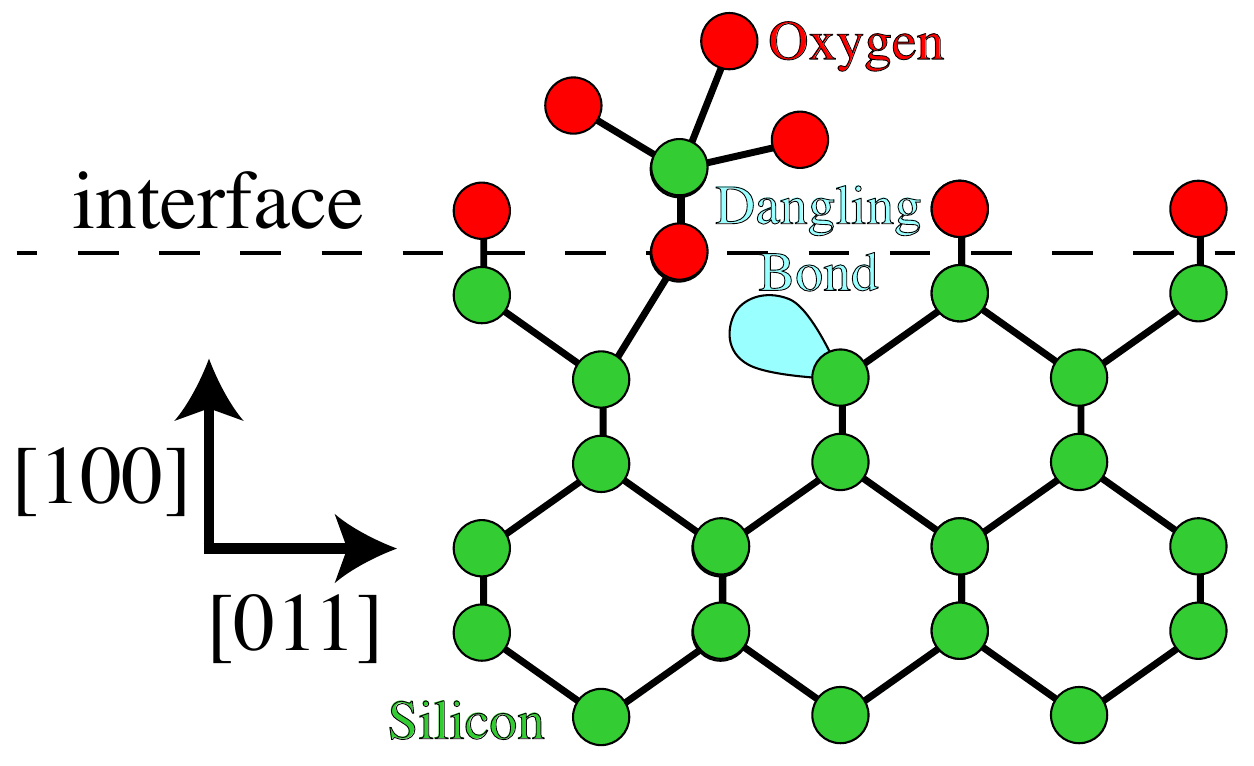}
        \caption[]
{The P$_{b0}$ defect at the (100) interface of Si and SiO$_2$. The defect is paramagnetic -- one unpaired electron is strongly localized at the trivalent back-bonded silicon atom. The P$_{b0}$ dangling bond symmetry axis is in the $\langle 111 \rangle$ family of directions.}\label{fig:Pb0} 
        \end{centering}
 \end{figure}
 \begin{figure*}[ptbh]
 \begin{centering}
        \includegraphics[scale = 0.385,trim = 0 10 0 5, angle = -0,clip]{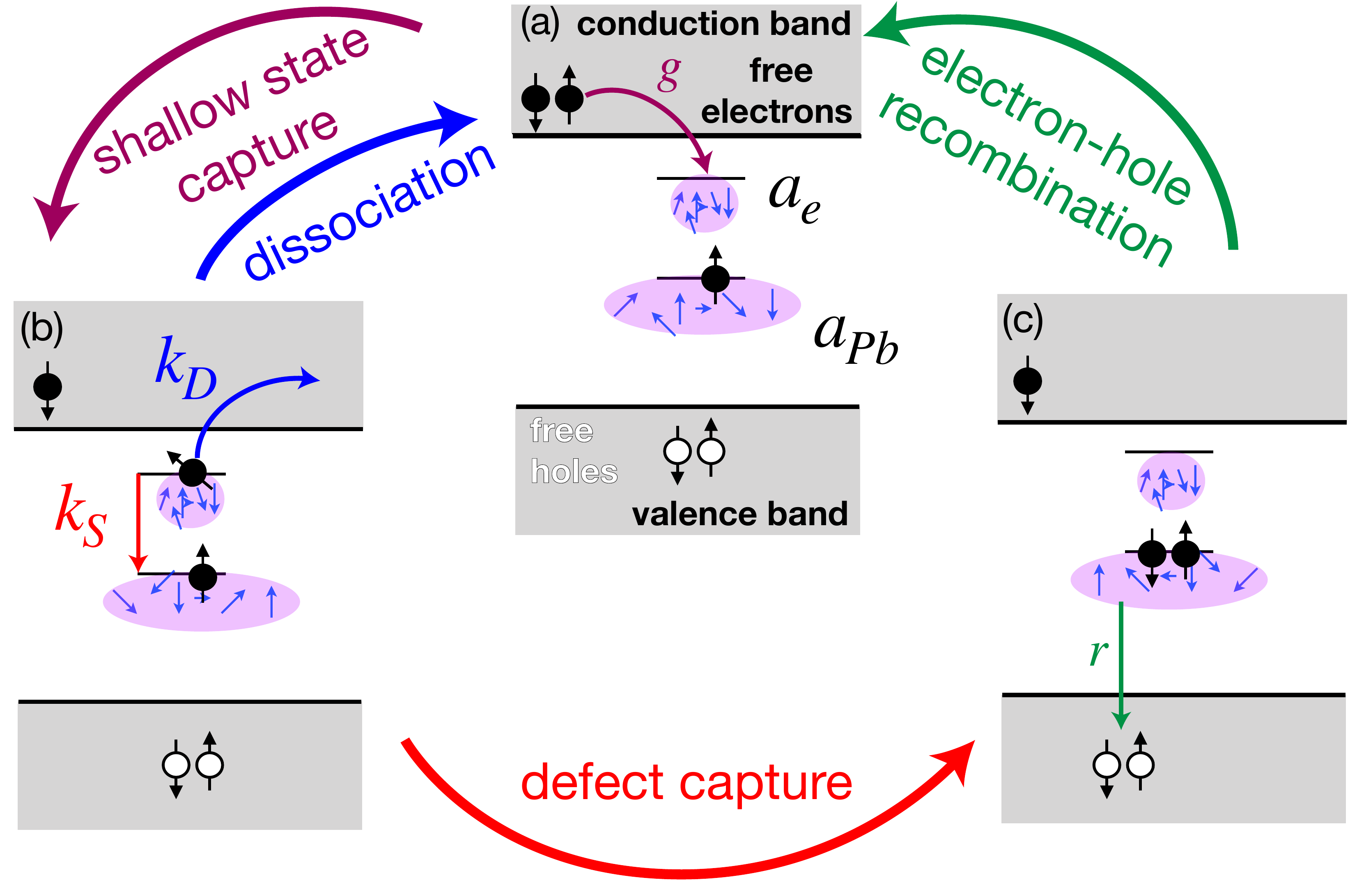}
        \caption[]
{Spin-dependent recombination at the Si/SiO$_2$ interface.  (a) Free electrons encounter a P$_b$ defect through a shallow state (\emph{e.g.} an excited state of the defect). (b) A weakly localized electron either dissociates at rate $k_D$ (return to (a)). Only a singlet state can exist at the defect site. If that is the case the shallow electron may be captured by the defect at rate $k_S$ (proceed to (c)). (c) An extra electron at the defect recombines with a free hole at rate $r$ which  returns the defect to its paramagnetic state (a). The blue arrows represent nuclear spins in the vicinity of the electron states, where $a_e$ and $a_{P_b}$ represent the hyperfine coupling constant for the shallow electron and P$_b$ defect, respectively.}\label{fig:RongModel} 
        \end{centering}
\end{figure*}

The high sensitivity of EDMR measurements can also make measurements requiring high sensitivity for other reasons considerably more straightforward; for example, Brower was first to report the hyperfine tensor components of  P$_{b}$ centers on the (111) SiO$_2$/Si interface utilizing conventional EPR; his measurements were close to the absolute sensitivity limits of conventional EPR, involving the stacking of 35 extensively thinned of Si/SiO$_2$ structures within an EPR cavity. In MOS technology, however, the (111) interface is almost never utilized; the (100) interface is almost universal, primarily because the density of interface trap defects (mostly the same P$_b$ centers) is significantly lower on the (100) interface. The geometry of the (100) interface also leads to multiple defect orientations. The lower density and somewhat more complex geometry make measurements of the hyperfine parameters of the dominating (100) P$_b$ variant, the P$_{b0}$ center (Fig. \ref{fig:Pb0}), significantly more difficult.  However such measurements become relatively straightforward with EDMR. Likely as a result of these sensitivity issues, the first measurements of the P$_{b0}$ hyperfine tensor components utilized EDMR.\cite{Gabrys1993}

EDMR measurements require the simultaneous presence of a large nearly static magnetic field and a radio or microwave frequency field. Because these oscillating fields are effectively shielded by conductive layers, the technique is untenable as a probe for defects at material interfaces within three dimensional integrated circuits.  EDMR is also unable to probe other material interfaces in which surrounding metallizations shield the oscillating field.

Recently spin-dependent recombination measurements displayed magnetoresistive effects near zero field whether an alternating field was applied or not. The magnitude of this near-zero field magnetoresistance (NZFMR) may be comparable to that of EDMR.\cite{Cochrane2012, Cochrane2013, Ashton2019} NZFMR was shown to be both present in a wide variety of systems and sensitive to radiation damage in those systems. The likely connection between NZFMR and the same defects that play a role in EDMR suggests that NZFMR spectroscopy could be a new tool to study defects in materials such as semiconductors and insulators, and especially in the very small interface regions of technologically-relevant MOS devices.\cite{Harmon2020}  The finer structure of the NZFMR line shapes even suggests that NZFMR may yield information not accessible through EDMR measurements.
NZFMR occurs due to correlations between at least two spins when the spins recombine to form either a singlet or triplet state. The Pauli exclusion principle dictates that $S=0$ and $S=1$ are nonequivalent; the simplest example is that if the spins are to lie in the same orbital level, the $S=1$ configuration is forbidden. The spin selection rules are important for recombination dynamics, the focus of this article, and also for understanding certain transport phenomena like trap-assisted tunneling\cite{Mott1987, Anders2018}, the subject of future work.
Similar spin correlated transport or luminescence has been studied in organic semiconductors over the past couple of decades.\cite{Mermer2005a, Prigodin2006, Macia2014, Wang2017}

This article provides the theoretical underpinnings of NZFMR spectroscopy for spin-dependent recombination at an oxide-semiconductor interface; this requires modifying the conventional Shockley Read Hall description\cite{Shockley1952, Hall1952} of trap-assisted recombination. Our approach utilizes a set of equations known as the stochastic Liouville equations which naturally account for the spin-selective processes. The formalism is used to determine line shapes of NZFMR and EDMR as a function of forward bias. In solids, the electronic $g$-factor of different electronic states may deviate from its established value of 2.002319 due to spin-orbit interactions. At large fields, the discrepancy in $g$-factors for different spin states is observable in EDMR. We focus specifically on applied static fields that are small enough that differences in effective $g$-factors between the two correlated spins do not play a role; this assumption allows us to explore solely the part played by the hyperfine interactions. We perform calculations of the line shape with two different models of the nuclear spin(s): semiclassical (spin vectors) and quantum mechanical (spin operators). We also examine the possibility that the captured carrier spin, in addition to the unpaired spin at the capturing defect site, experiences a hyperfine interaction from nearby atoms. Si/SiO$_2$ MOSFETs are used as a case study to compare with calculations. By doing so, we are able to obtain a detailed understanding of the physics of spin-dependent recombination for Si/SiO$_2$ MOSFET system. The results from our modeling suggest that NZFMR spectroscopy may serve as a new diagnostic for defects in semiconductors and insulators, and especially at the interfaces between them.

\section{Spin-Dependent Recombination} 

In the early 1970s, Lepine discovered that spin-dependent recombination would lead to changes in conductivity in silicon. Lepine's analysis contended that, in the presence of a magnetic field, recombining spins were thermally polarized and less likely to recombine based on the Pauli exclusion principle \cite{Lepine1972a, Lepine1972b}. The theoretical predictions of this polarization model were woefully unmet -- the relative recombination rates were much larger than predicted and carried little magnetic field or temperature dependence. A variety of polarization models failed to improve the situation \cite{Lvov1977, Wosinski1977, Mendz1980}.

In 1978 an advance occurred when Kaplan, Solomon, and Mott (KSM) produced a new model not based on spin polarization \cite{Kaplan1978}. This KSM model posited that spins underwent an intermediary phase before recombination. Once the pair enters this intermediary phase (which might be an exciton or a donor-acceptor pair), the pair components are \emph{exclusive} to one another; the pair has the option to either recombine or dissociate. The recombination process is spin-dependent while dissociation is not. Only after dissociation can either component interact, and possibly recombine, with other carriers.
The KSM model explained many of the difficulties that confronted the Lepine model.
Further adaptation was provided by Rong \emph{et al.} where the intermediate state was supposed to be either an excited state of the defect or a shallow donor state near the conduction band.\cite{Rong1991} The carrier is first trapped into this excited state where it then has some probability of either dissociating back into an itinerant state or falling down into the ground state of the defect.
It is this last process which is spin-dependent due to the Pauli exclusion principle since the defect is initially paramagnetic. The singlet ground state of the charged defect may capture and recombine with a hole and the process repeats with the defect becoming paramagnetic once more.\cite{Spaeth2003} These processes are depicted in Figure \ref{fig:RongModel}.

In this article we use the idea of exclusive spin pairs to model both EDMR and NZFMR. Our approach utilizes stochastic Liouville equations for a spin density matrix which allows for a more general treatment than either the theories of KSM or Rong provided in the past. Previous theories of EDMR have been limited in scope by treating only transitions induced by different $g$-factors or semiclassical nuclear fields.\cite{Glenn2013a, Glenn2013b, Limes2013} When examining NZFMR in systems with very few nuclear spin, it is imperative to include fully quantum hyperfine interactions. In the article we analyze NZFMR and EDMR by treating both quantum and semi-classical hyperfine interactions and then discuss the appropriate approach to use for our Si/SiO$_2$ MOSFET devices. In the next section we alter the conventional model of trap-assisted recombination to include the ideas of pair exclusivity from KSM and the spin dependence of carrier trapping at a deep level paramagnetic defect site.

\begin{figure*}[ptbh]
 \begin{centering}
        \includegraphics[scale = 0.8,trim = 0 505 0 80, angle = -0,clip]{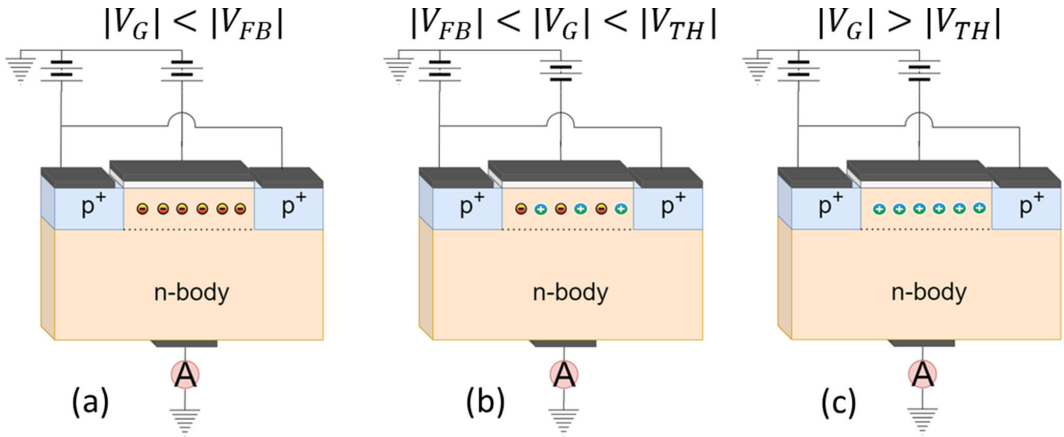}
        \caption[]
{Schematic illustration of the gated-diode biasing scheme. (a) The MOSFET is in accumulation and only electrons occupy the interface region. (b) The MOSFET is in depletion and, assuming the densities of electrons and holes are equal, interface recombination is maximized. (c) The MOSFET is in inversion and the interface channel region now consists of mostly holes. }\label{fig:experiment} 
        \end{centering}
\end{figure*}

\section{Spin-Dependent Recombination Current}

Recombination current in a semiconductor device can be explained utilizing the Shockley-Read-Hall model of recombination. In this model, electron-hole pair recombination most effectively takes place at deep level defect centers near mid-bandgap.\cite{Shockley1952, Hall1952} Thus, in measurements of interface recombination utilizing a gated diode measurement, the recombination defect centers have energy levels very near the middle of the Si bandgap. Recombination current results from capture of both types of charge carriers at a deep level. Consider the simple case of an electron traveling through the conduction band. When the electron encounters a deep level defect, it may fall into the deep level granted that spin selection rules are obeyed. Once the electron capture takes place, the electron is available for recombination with a hole in the valence band. This process may also take place via hole capture and subsequent recombination with a conduction band electron. In the gated diode measurement, the recombination current measured through the body contact of the MOSFET has a peak which corresponds to the situation where both electron and hole densities are equal. When this occurs the recombination is maximized. 

The peak of recombination current can be thought of quantitatively as follows. 
Under conditions where the MOSFET's source and drain are shorted together under a forward bias with gate voltage set to yield maximum interfacial trap recombination current (see Fig. \ref{fig:experiment} which is configuration of experiment used here), the recombination current is
\begin{equation}\label{eq:Imax0}
I_{max} = \frac{1}{2} q^2 n_i \sigma v_{th} A D_{it} |V_F| e^{q|V_F|/2 kT}
\end{equation}
where $\sigma$ is the equal electron and hole capture cross sections by the defect, $v_{th}$ is the thermal velocity of the carriers, $n_i$ is the intrinsic carrier concentration, $D_{it}$ is the areal concentration of interface traps per energy, $A$ is the gate area, $q$ is the electronic charge, $V_F$ is the forward bias voltage, $k$ is Boltzmann's constant, and $T$ is the absolute temperature. This $I_{max}$ is derived independent of spin.\cite{Fitzgerald1968} 
In the experiments conducted here (to be discussed further in Section \ref{sec:experiment}), under forward bias holes of concentration $p$ are injected into the channel. If $p_i \equiv n_i$, then for room temperature Si $n_i$ is about 1.5 $\times 10^{10}$ cm$^{-3}$,  and with forward bias, $p = n_i e^{q |V_F|/2k_B T}$ which gives a range of carriers between about 2 $\times 10^{15}$ cm$^{-3}$ and 7 $\times 10^{18}$ cm$^{-3}$ for the range of forward biases used here.

Equation (\ref{eq:Imax0}) is insufficient for a few reasons: the electron and hole cross sections, $\sigma_e$ and $\sigma_h$, are unequal\cite{Garrett1956}; the recombination is mediated by an intermediate shallow state; the capture of an electron by the deep trap is spin dependent. 
Equation (\ref{eq:Imax0}) can be modified to include these additional effects as described next.

\subsection{Calculation for maximum spin-dependent recombination current}

The maximum recombination current is
\begin{equation}
I_{max} = q A U_s
\end{equation}
where $A$ is the gate area so $U_s$ has units of inverse time and area. The current is maximum when the gate bias is tuned in a way that there are equal numbers of electrons and holes at the interface which maximizes $U_s$.

The spin-dependent capture of a carrier electron by a deep trap is a two step process: (1) the carrier electron is first weakly localized by a shallow state in the vicinity of the defect. At room temperature this state is most likely an excited state of the defect which may be near the conduction band.\cite{Rong1991, Boehme2004, Friedrich2005, Hori2019} We assume this going forward.
(2) the electron in the now charged excited state of the deep paramagnetic trap reduces to its charged ground state if the singlet condition for the spin pair is met. The theory of Shockley, Read\cite{Shockley1952},  and Hall\cite{Hall1952} can be modified to include the two-step capture (into the ground state) by the trap. The rate of capture of a conduction electron into the trap ground state is $c_n N_T$ where we call $c_n$ the capture parameter (with dimension rate $\times$ volume) and  $N_T$ is the density of traps. If the two steps are each accomplished with respective capture parameters $c_{t^*}$ and $c_t$ then the 
total capture parameter of a conduction electron into the deep level trap by way of its excited state is
\begin{equation}
c_{n}  = \frac{c_{t^*} c_t}{c_{t^*} + c_t} \approx  c_t.
\end{equation}
The last approximation is made since the transition rate to the trap ground state (large energy difference) is much smaller than the capture of a conduction electron by the excited state (small energy difference).
The capture rate per volume accounts for the concentration of conduction electrons, $n$, and is
\begin{equation}
r_t = c_t n N_T (1 - f(E_t))
\end{equation}
where $f(E)$ is the non-equilibrium occupancy factor at the trap level which can be determined in the steady state.\cite{Shockley1952, Hall1952}
The calculation can then proceed in a straightforward manner, following that of Shockley, Read\cite{Shockley1952},  Hall\cite{Hall1952},  Fitzgerald and Grove\cite{Fitzgerald1968}.
These calculations assume that traps have constant density throughout gap but the recombination process is dominated by traps near the center. \footnote{The calculation can also be done with a single defect level at mid gap. The purpose here is to mimic the Fitzgerald and Grove model since that is the one that gets quoted most often in the literature.}

The recombination rate per unit area is:
\begin{widetext}
\begin{equation}\label{eq:US1}
U_s = c_{n} \sigma_p v_{th} D_{st} \left[  \int_{E_v}^{E_c} \frac{dE_{st}}{c_{n}( n_s +  n_i e^{(E_{st} - E_i)/k_B T})  + \sigma_p v_{th}(p_s  +  n_i e^{-(E_{st} - E_i)/k_B T})} \right] (p_s n_s - n_i^2)
\end{equation}
\end{widetext}
where $E_{st}$ is the energy of the recombination center and $D_{st}$ is the areal density per energy of such centers. The quantities $n_s $ and $p_s$ have dimensions of inverse volume. 
As mentioned earlier, the deep defects possess the relevant capture parameter $c_n = c_t$ which, unlike for holes, should not be expressed as $\sigma_n v_{th}$ because the electron is already situated at the defect. 
Since relaxation of the excited electron spin into the ground state of the defect is spin dependent, $c_{n} \approx (k_S \rho_S/g) c_{n,0}$ where $c_{n,0}$ is the maximum possible capture parameter and $(k_S \rho_S/g)$ is the probability of a singlet that comes from the stochastic Liouville equation which is demonstrated in the next section.

This integral in Eq. (\ref{eq:US1}) can be determined to be approximately\footnote{The intrinsic Fermi level, $E_i$, is assumed to be at mid gap such that $(E_v - E_i )/ k_B T \ll 0$ and $(E_c - E_i )/ k_B T \gg 0$ }
\begin{equation}
U_s = \sqrt{c_n \sigma_p v_{th}}  k_B T D_{st} \frac{\text{arccosh}(x)}{n_i\sqrt{x^2 - 1}}(p_s n_s - n_i^2)
\end{equation}
with 
\begin{equation}
x = \frac{p_s}{2 n_i}\sqrt{\frac{\sigma_p v_{th}}{c_n}}  + \frac{n_s}{2 n_i}\sqrt{\frac{c_n}{\sigma_p v_{th}}} .
\end{equation}
The function $\text{arccosh}(x)/\sqrt{x^2 - 1}$ decays monotonically from a maximum value of $\pi/2$. Thus maximizing the recombination rate entails minimizing the quantity $x$. Given that $n_i$ is the intrinsic carrier density, the only way to do so is by minimizing both $n_s$ and $p_s$. 
Just as for Fitzgerald and Grove, the minimum values for these two are $n_i e^{q |V_F|/2k_B T}$ which can be reached by tuning the gate voltage. Note that the unequal cross sections do not change this criterion.

Substituting in these constraints on the electron and hole densities yields
\begin{widetext}
\begin{equation}
U_s = \sqrt{c_n \sigma_p v_{th}}  k_B T D_{st} \frac{\text{arccosh}\left[\frac{e^{q |V_F|/2k_B T}}{2}(\sqrt{\frac{\sigma_p v_{th}}{c_n}}  + \sqrt{\frac{c_n}{\sigma_p v_{th}}} )\right]}{\sqrt{\frac{e^{q |V_F|/k_B T}}{4}(\sqrt{\frac{\sigma_p v_{th}}{c_n}}  + \sqrt{\frac{c_n}{\sigma_p v_{th}}} )^2 - 1}}(e^{q |V_F|/k_B T} - 1)n_i
\end{equation}
In the limit of large forward bias, the result simplifies to 
\begin{equation}
U_s = 2 \sqrt{c_n \sigma_p v_{th}}  k_B T D_{st} \frac{\text{arccosh}\left[\frac{e^{q |V_F|/2k_B T}}{2}(\sqrt{\frac{\sigma_p v_{th}}{c_n}}  + \sqrt{\frac{c_n}{\sigma_p v_{th}}} )\right]}{ \sqrt{2+ \frac{\sigma_p v_{th}}{c_n} +\frac{c_n}{\sigma_p v_{th}} } }   e^{q |V_F|/2 k_B T} n_i
\end{equation}
 \end{widetext}
The argument of $\text{arccosh}$ is large and so can be expanded as $\text{arccosh}(y) \approx \ln(2) - \ln (y^{-1})= \ln(2) + \ln (y) =  \ln(2) + \ln (1/2) + \ln(\sqrt{\frac{\sigma_p v_{th}}{c_n}} + \sqrt{\frac{c_n}{\sigma_p v_{th}}}) + \frac{q |V_F|}{2 k_B T} $ where only the last term is appreciable under physical conditions.
This finally yields
\begin{equation}
U_s = \frac{ \sqrt{\sigma_p v_{th} c_{n}}}{\sqrt{2+\frac{\sigma_p v_{th}}{c_n}  + \frac{c_{n}}{\sigma_p v_{th}} }}    D_{st}q n_i |V_F| e^{q |V_F|/2 k_B T} .
\end{equation}
 An effective cross-section can be defined as: 
 \begin{equation}\label{eq:Sigma}
\Sigma = 2\frac{ \sqrt{\sigma_p v_{th} c_{n}}}{\sqrt{2+\frac{\sigma_p v_{th}}{c_{n}}  + \frac{c_{n}}{\sigma_p v_{th}} }} 
\end{equation}
such that
\begin{equation}
U_s = \frac{1}{2}\Sigma  D_{st}q n_i |V_F| e^{q |V_F|/2 k_B T} .
\end{equation}
In the case of equal capture rates ($\sigma v_{th}$), 
\begin{equation}
U_s = \frac{ \sigma}{2}  v_{th}  D_{st}q n_i  |V_F| e^{q |V_F|/2 k_B T} 
\end{equation}
which is exactly the result of Fitzgerald and Grove.


\subsection{The effective recombination cross section}

\begin{figure}[ptbh]
 \begin{centering}
        \includegraphics[scale = 0.35,trim = 0 405 420 0, angle = -0,clip]{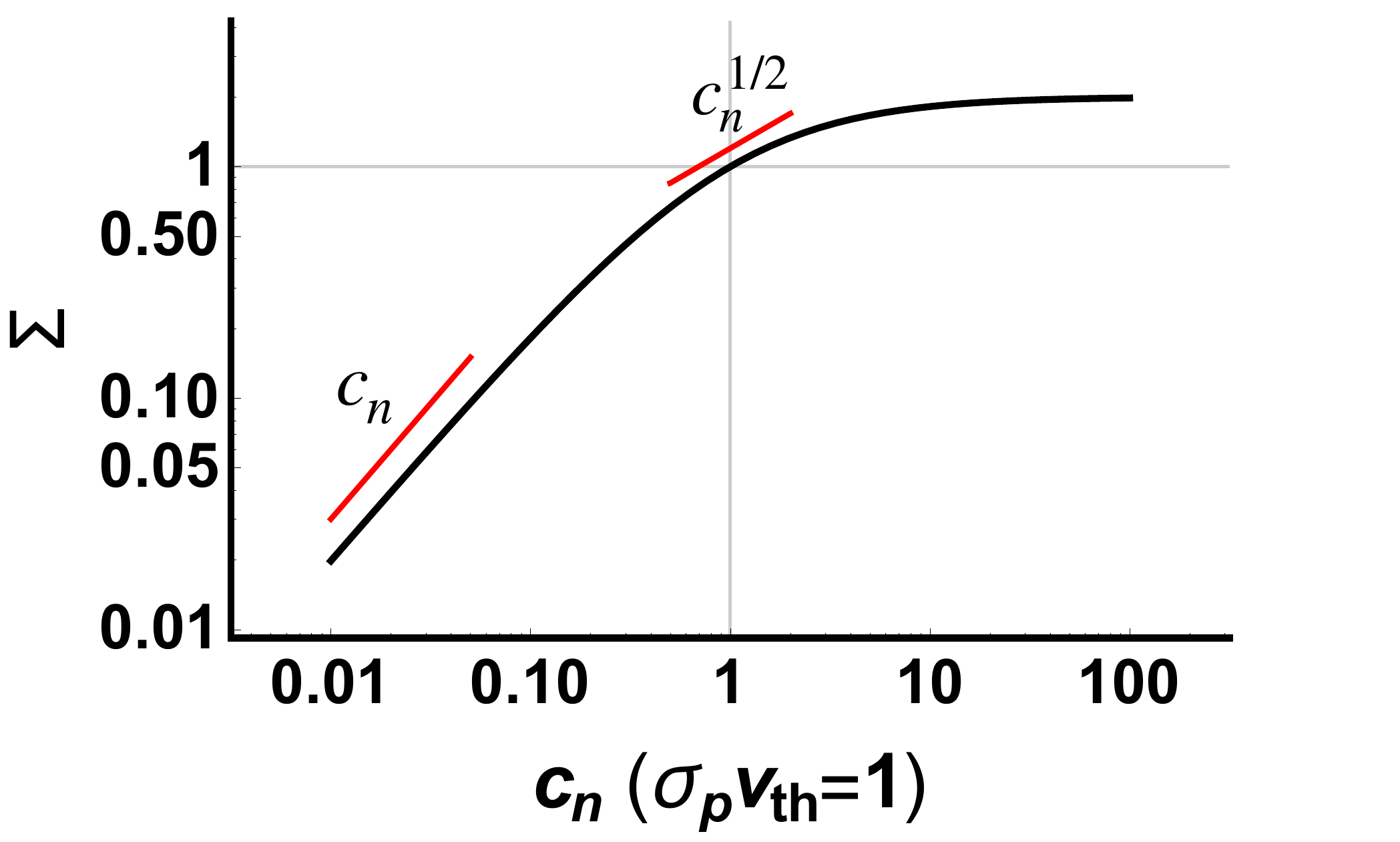}
        \caption[]
{Effective cross section as function of electron capture parameter. Line segments depict regions of $\sim c_n^1$ and $\sim c_n^{1/2}$}\label{fig:crosssection} 
        \end{centering}
\end{figure}

The spin dependence of the recombination enters through the effective cross section, $\Sigma$, in Eq. \ref{eq:Sigma}.
Fig. \ref{fig:crosssection} plots effective cross section versus electron capture rate per area of the trap. For small $c_n$ compared to $\sigma_p v_{th}$, the dependence is linear.  We can think about it in this way: when the holes are very rapidly captured, electron capture is the limiting step so any increase in recombination requires quicker electron capture. The singlet probability, $\rho_S$, enters within $c_n \propto \rho_S$.
This calculation assumes the cross sections or capture parameters are independent of energy. There are interpretations of the cross section that might suggest that assumption to be incorrect.\cite{Ryan2015}
The calculation would obviously be more difficult and probably not allow for any analytic solution. 

Our system of a P$_{b0}$ defect at the Si/SiO$_2$ interface is clearly in the regime of $c_n \ll \sigma_p v_{th}$ since the trap will be negatively charged when capturing a hole but neutral when capturing an electron.  So our final expression for the maximum recombination current is
\begin{equation}\label{eq:maxCurrentFinal}
I_{max} = q^2 A c_n  D_{st} n_i |V_F| e^{q |V_F|/2 k_B T}  = q^2 A \frac{k_S \rho_S}{G}  c_{n,0} D_{st} n_i |V_F| e^{q |V_F|/2 k_B T} .
\end{equation}
This expression is valid even when a magnetic field is incuded (through the parameter $\rho_S$  -- see next section) as long as the bias condition for maximum current is unchanged by the magnetic field. The optimum current position has not been systematically studied in relation to magnetic field but one study shows very little change in the bias condition. \cite{Campbell2007} 
Therefore Eq. (\ref{eq:maxCurrentFinal}) applies for the NZFMR and EDMR experiments described in this article.

All other quantities in $I_{max}$ are considered to be spin or magnetic field-independent; hence $I_{max}(B_0)/I_{max}(\infty)  = c_n(B_0)/c_n(\infty) = \rho_S(B_0)/\rho_S(\infty)$ is the normalized current at a given forward bias.

\section{The Stochastic Louiville Equation and the Lyapunov Equation}

We use a spin-density matrix $\rho$ which fully accounts for not only the spin-pair but also any number of nuclear spins. The evolution of the spin pair plus any number of relevant nuclei is governed by the stochatic Liouville equation:
\begin{widetext}
\begin{equation}\label{eq:sle}
\frac{\partial \rho}{\partial t} = -\frac{i}{\hbar} [\mathscr{H}_{0}+ \mathscr{H}_{hf} + \mathscr{H}_{1}  , \rho] - \frac{k_S+k_D}{2} \{ P_S, \rho \}  - \frac{k_T + k_D}{2} \{ P_T, \rho \}  +  \frac{g}{\text{Tr}\mathbb{1}} \mathbb{1},
\end{equation}
\end{widetext}
where we use SI units and the Hamiltonians are presented in detail in the next section. The first term on the right-hand side is the Liouville or Neumann equation for the density matrix, describing the coherent evolution of the density matrix.  The second and third terms signify the random processes of spin capture and spin dissociation of the spin pairs which in general may depend on their spin configuration (singlet or triplet combination occur at rates $k_S$ and $k_T$, respectively). Assuming small spin-orbit interactions, we take $k_T = 0$; triplets are not captured by the deep defect. The rate of singlet capture depends on the occupation of states so is written as $k_S \text{Tr}[P_S \rho(t)] = k_S \rho_S(t)$. A rate $k_D$ describes the dissociation of the spin pair.

Consider the steady state $\frac{\partial \rho}{\partial t}  = 0$ which yields the following adaptation to Eq. \ref{eq:sle}:
\begin{equation}
\mathscr{L} \rho+ \rho \mathscr{L}^{\dagger}  = \frac{g}{\text{Tr}\mathbb{1}} \mathbb{1}.
\end{equation}
with
\begin{equation}
\mathscr{L} = \frac{i}{\hbar} \mathscr{H} + \frac{k_S+k_D}{2}P_S +\frac{k_D}{2} P_T.
\end{equation}
This type of type of equation is known as a Lyapunov equation.
$\mathscr{L} $ can also be written as an effective Hamiltonian $H = - i \hbar \mathscr{L} $.
The solution for the Lyapunov equation is then
\begin{equation}\label{eq:rhoSol}
\rho = \frac{g}{\text{Tr}\mathbb{1}} \int_0^{\infty}e^{-i H t }e^{i H^{\dagger} t} dt =  \frac{g}{\text{Tr}\mathbb{1}}  U
\end{equation}
where
\begin{equation}
U = \int_0^{\infty}e^{-i H t }e^{i H^{\dagger} t} dt.
\end{equation}

Another relation is obtained by taking the trace of Eq. \ref{eq:sle} in the steady state:
\begin{equation}\label{eq:Tracesle}
- (k_S+k_D) \text{Tr}( P_S \rho) - k_D  \text{Tr}( P_T \rho)   + g =    0 ,
\end{equation}
which leads to 
\begin{equation}
\frac{ k_S \text{Tr}( P_S \rho)}{g}  + \frac{k_D  \text{Tr} \rho   }{g} = 1.
\end{equation}
The first term is the fractional yield of captured singlet pairs (creating the $S =0$ defect state); the second term is the fractional yield of dissociated spin pairs. 
Another way to write this is
\begin{equation}
k_S \rho_S \equiv k_S \text{Tr}( P_S \rho) = g - k_D  \text{Tr} \rho  
\end{equation}
which is combined with Eq.\ref{eq:rhoSol} to ascertain
\begin{equation}\label{eq:eq23}
k_S \rho_S =   g  - k_D  \text{Tr} (\frac{g}{\text{Tr}\mathbb{1}}  U)  = g  \left (1-  \frac{k_D}{\text{Tr}\mathbb{1}}  \text{Tr} (U)\right) .
\end{equation}
Spin relaxation and decoherence are ignored in Eq. (\ref{eq:eq23}).

The solution to the SLE gives
\begin{equation}
\rho_S = \frac{g}{\text{Tr}\mathbb{1}} f(B_0)
\end{equation}
where $\text{Tr}\mathbb{1} = 2^n$ is the spin degeneracy factor ($2^n$ where $n$ is number of spins included in the density matrix; $2^n$ is also the dimension of the density matrix here).
$f(B_0)$ is the remaining portion of the density matrix. $g$ is the rate that which spin pairs are generated. So this factor depends on the e-h recombination and the capture of an electron by the shallow state. If we assume that e-h recombination occurs rapidly after the defect becomes $S= 0$, then the rate $g$ is limited by the shallow state's capture of electrons. Therefore we write $g = \sigma_{t^*} v_{th} N_{dt}$ where $\sigma_{t^*}$ is the capture cross section of the shallow state (assumed to be excited level of the trap), $v_{th}$ is the thermal velocity of conduction 
electrons, and $N_{t}$ is the density of traps.

\section{Interactions}

\subsection{Hyperfine Interaction at Defect}

We ignore differences and anisotropies in carrier or defect g-factors by assuming that the radio or microwave frequency is small enough such that large static magnetic fields are unnecessary to probe the EDMR dynamics. We start by assuming only one of the two non-nuclear spins is interacting with any number of nuclear spins:
\begin{eqnarray}
&&\mathscr{H}_{0} = g_e \mu_B B_0 (S_{z,1} + S_{z,2}) -  g_n \mu_n B_0 \sum_j I_{z, j},{}\\ {}&&\mathscr{H}_{hf} = g_e \mu_B  \sum_j \bm{I}_j\cdot \hat{\bm{A}}_j \cdot \bm{S}_1,{}\\
{}&&\mathscr{H}_{1} = 2 B_1 ( g_e \mu_B S_{x,1} +  g_e \mu_B S_{x,2} -  g_n \mu_n \sum_j I_{x, j}) \cos\Omega t.
\end{eqnarray}
The first Hamiltonian is the static Zeeman interaction for the electrons and nuclei. The second Hamiltonian is the hyperfine interaction, and the third Hamiltonian is the interaction of the spins with the transverse oscillating field. 
After making the assumption that $g_n \mu_n \ll g_e \mu_B$, the time-dependent Hamiltonian can be decomposed into two pieces $\mathscr{H}_{1} = \mathscr{H}_{a}+\mathscr{H}_{b}$:
\begin{widetext}
\begin{equation}
\mathscr{H}_{a} = B_1 ( g_e \mu_B S_{x,1} +  g_e \mu_B S_{x,2}) \cos(-\Omega t )+ B_1  ( g_e \mu_B S_{y,1} +   g_e \mu_B S_{y,2} )\sin(-\Omega t),
\end{equation}
and 
\begin{equation}
\mathscr{H}_{b} = B_1 ( g_e \mu_B S_{x,1} +  g_e \mu_B S_{x,2} ) \cos\Omega t + B_1  ( g_e \mu_B S_{y,1} +   g_e \mu_B S_{y,2} )\sin\Omega t.
\end{equation}
The rotating wave approximation (RWA) leads to dropping $\mathscr{H}_{b}$ (the details are described in the Appendix):
\begin{equation}
\tilde{\mathscr{H}}_{} = g_e \mu_B  (B_0 + \frac{\hbar \Omega}{g_e \mu_B }) (S_{z,1} + S_{z,2}) + \hbar \Omega \sum_j I_{z, j} +  g_e \mu_B  \sum_j \bm{I}_j\cdot \hat{\bold{A}}_j \cdot \bm{S}_1 +  g_e \mu_B B_1 (S_{x,1} + S_{x,2})
\end{equation}
\end{widetext}
where we do need to assume that the hyperfine interactions have the same principle axes and possess axial symmetry.
The Liouville equation becomes
\begin{equation}
\frac{\partial \tilde{\rho}}{\partial t} = -\frac{i}{\hbar} [\tilde{\mathscr{H}}_{}, \tilde{\rho}] - \frac{k_S+k_D}{2} \{ P_S, \tilde{\rho} \}  - \frac{k_D}{2} \{ P_T, \tilde{\rho} \}+  \frac{g}{\text{Tr}\mathbb{1}}\mathbb{1}
\end{equation}
where $\tilde{\rho}$ is the density matrix in the rotating reference frame.
It is important to note that the Hamiltonians in the rotating frame are time independent.

\subsection{Hyperfine Interactions at Both Electron Spins}

The derivation carried out in the previous section can be repeated exactly when there is also a hyperfine interaction at $\bold{S}_2$. The total hyperfine interaction is then:
\begin{equation}
    \tilde{\mathscr{H}}_{hf} = g_e \mu_B  \sum_j \bm{I}_j\cdot \hat{\bold{A}}_j \cdot \bm{S}_1 + g_e \mu_B  \sum_j \bm{I}_j\cdot \hat{\bold{B}}_j \cdot \bm{S}_2
\end{equation}
in the rotating wave approximation  with the same aforementioned restrictions on both hyperfine interactions.

\subsection{The Semiclassical Approximation}

In the limit of a large number of nuclear spins composing the hyperfine interactions at either or both of the sites for $\bm{S}_1$ and $\bm{S}_2$, the quantum mechanical nuclear spin operator is replaced by a classical vector which physically corresponds to an ensemble of nuclear moments interacting with the single electron spin \cite{Flory1969,Schulten1978,  Rodgers2007}:
\begin{equation}
\mathscr{H}_{hf} = g_e \mu_B  \bm{B}_{n,1} \cdot \bm{S}_1 +  g_e \mu_B  \bm{B}_{n,2} \cdot \bm{S}_2
\end{equation}
where the probability distribution functions for the nuclear fields pointing at random angles are
\begin{eqnarray}\label{eq:gaussian}
    W(\bm{B}_{n,1}) = \left(\frac{1}{2 \pi  a^2}\right)^{3/2} e^{-\frac{B_{n,1}^2}{2 a^2}},{}\\ ~~~~W(\bm{B}_{n,2}) = \left(\frac{1}{2 \pi b^2} \right)^{3/2} e^{-\frac{B_{n,2}^2}{2  b^2}},
\end{eqnarray}
with effective hyperfine fields $a^2 = \frac{1}{3} \sum_j a_j^2 I_j (I_j + 1)$ and $b^2 = \frac{1}{3} \sum_j b_j^2 J_j (J_j + 1)$ with $I_i$ and $J_i$ being the spin quantum numbers for each of the nuclei at site one and site two, respectively. 
For simplicity, we have assumed the semiclassical distribution of fields to be isotropic (\emph{i.e.} $B_{n,1, x}$ has same root mean square field as $B_{n,1, z}$).

Unfortunately, the transformation of the semiclassical hyperfine Hamiltonian to the rotating frame does not yield a  Hamiltonian independent of time. Of course, this is inconsequential for NZFMR as we calculate with no RF field. 

If one makes the secular approximation to the hyperfine interaction (which is valid for large field), then the rotating wave approximation can be used to obtain
\begin{eqnarray}
&& \tilde{\mathscr{H}}_{0} = g_e \mu_B  (B_0 + \frac{\hbar \Omega}{g_e \mu_B }) (S_{z,1} + S_{z,2}) ,{}\\ &&\tilde{\mathscr{H}}_{hf} = g_e \mu_B  B_{n,1, z}  S_{1,z} + g_e \mu_B  B_{n,2, z}  S_{2,z},{}\\
&&\tilde{\mathscr{H}}_{1} = g_e \mu_B B_1 (S_{x,1} + S_{x,2})
\end{eqnarray}
So EDMR and NZFMR calculations can still be satisfactorily carried out as long as $\hbar \omega/ g_e \mu_B \gg B_{n}$ (use secular approximation) or $B_1 = 0$ (no RF field at all).

\section{Hyperfine MC Line Shape Structure}

\subsection{Quantum Mechanical Nuclear Spins}

\begin{figure*}[ptbh]
 \begin{centering}
        \includegraphics[scale = 0.4,trim = 5 0 20 30, angle = -0,clip]{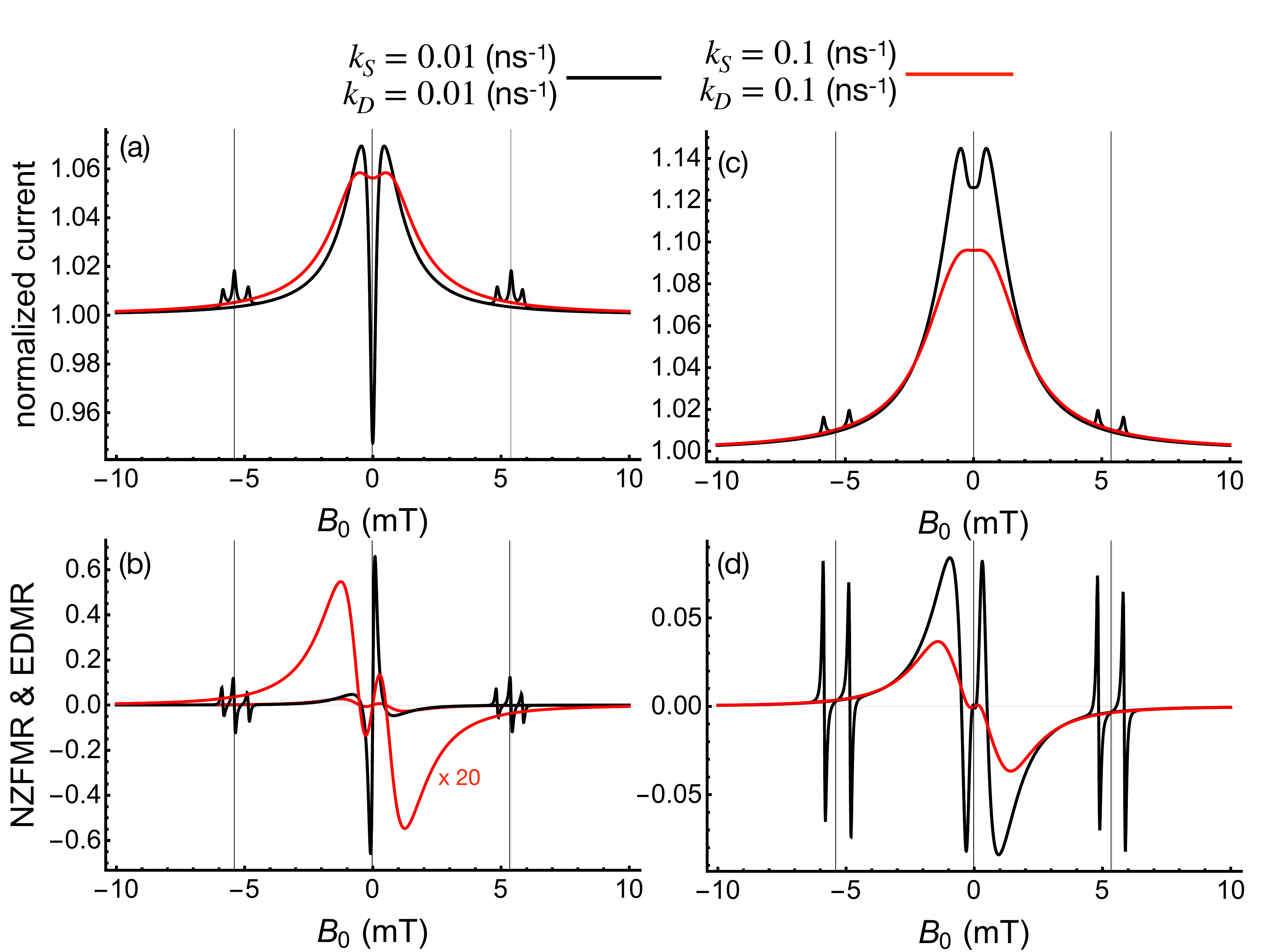}
        \caption[]
{Quantum calculation. (a,b) Two contrasting line shapes using a single spin-1/2 quantum nucleus at $\bm{S}_1$.
(c,d) Two contrasting line shapes using a single spin-1/2 quantum nucleus at $\bm{S}_1$ and a single spin-1/2 quantum nucleus at $\bm{S}_2$. Both calculations used $a_1 = 1$ mT, $g = 2$, $B_1 = 0.02$ mT, $ f= 150$ MHz ($\approx 5.4$ mT). (a,b) $a_2 = 0$ (c,d) $a_2 = 1$ mT. The RWA is used in each calculation.}\label{fig:QMlines} 
        \end{centering}
\end{figure*}

The recombination currents, either with or without the oscillating field, are guaranteed to have the same value only for large magnetic fields (\emph{i.e.} far from resonances and near zero field transitions); it is convenient then to use $\rho_S(B_0)/\rho_S(\infty)$ as a figure of merit. 
The most sensitive experiments determine  a spin-dependent recombination current which will be proportional to $\frac{d}{dB_0}\rho_S(B_0)/\rho_S(\infty)$ which is the quantity we call ``NZFMR" or ``EDMR".
Figure \ref{fig:QMlines} shows two representative calculations using different rates $k_S$ and $k_D$ with an isotropic hyperfine interaction at one (a,b) or both (c,d) of the sites.

The EDMR is nearly absent when the spin kinetics are increased (red) since the spins are either combining or dissociating before the resonant transition can take place. Similarly, the increased rates also cause an overall broadening and reduction in magnitude of the curve.

Note that in panels (a,c), the NZFMR line shape consists of a broad feature (`shoulder') and a narrow feature (`dimple'). The `dimple' appears more pronounced in the response derivative (b,d). The `shoulder' is barely observable in the black curve of (b) while the red curve's `dimple' in (b) is weaker than the `shoulder'. The black `shoulder', `dimple', and EDMR are comparable in (c) though reducing the `dimple' (by appropriate choice of parameters) tends to also result in a reduction of the EDMR as shown by the red curve. 
The `dimple' appears in fields less than $a$ and are similar to what has been observed in the spin chemistry of radical pair reactions (low-field effect or LFE) and the magnetic field effects studied in organic spintronics.
The origin is subtle and qualitatively ascribed to the conservation of angular momentum \cite{Brocklehurst1996}. 
Consider initial singlet spin pair states with a spin-$\frac{1}{2}$ nucleus located at one of the two spins.
In zero applied field  both $J^2$ and $J_z$ are conserved where $\bold{J} = \bold{S} + \bold{I}$. The conservation of these quantities restricts the overall hyperfine-induced transitions between singlet and triplet states. 
When a small field is applied (along $\hat{z}$) only $J_z$  is conserved. As a result the spin evolution is given more freedom when this small field is applied and more triplets may be formed at the expense of the singlet population.
As the field increases, eventually energetics play a role and the non-zero $m_s$ states separate out from the $m_s = 0$ states which gives rise to fewer transitions out of the singlet state and an eventual saturation of the singlet population. 

We observe here, and we find this to be true in general, that the `dimple' originating from the LFE is typically smaller in size when hyperfine interactions exist at both sites.

\begin{figure*}[ptbh]
 \begin{centering}
        \includegraphics[scale = 0.5,trim = 0 0 20 10, angle = -0,clip]{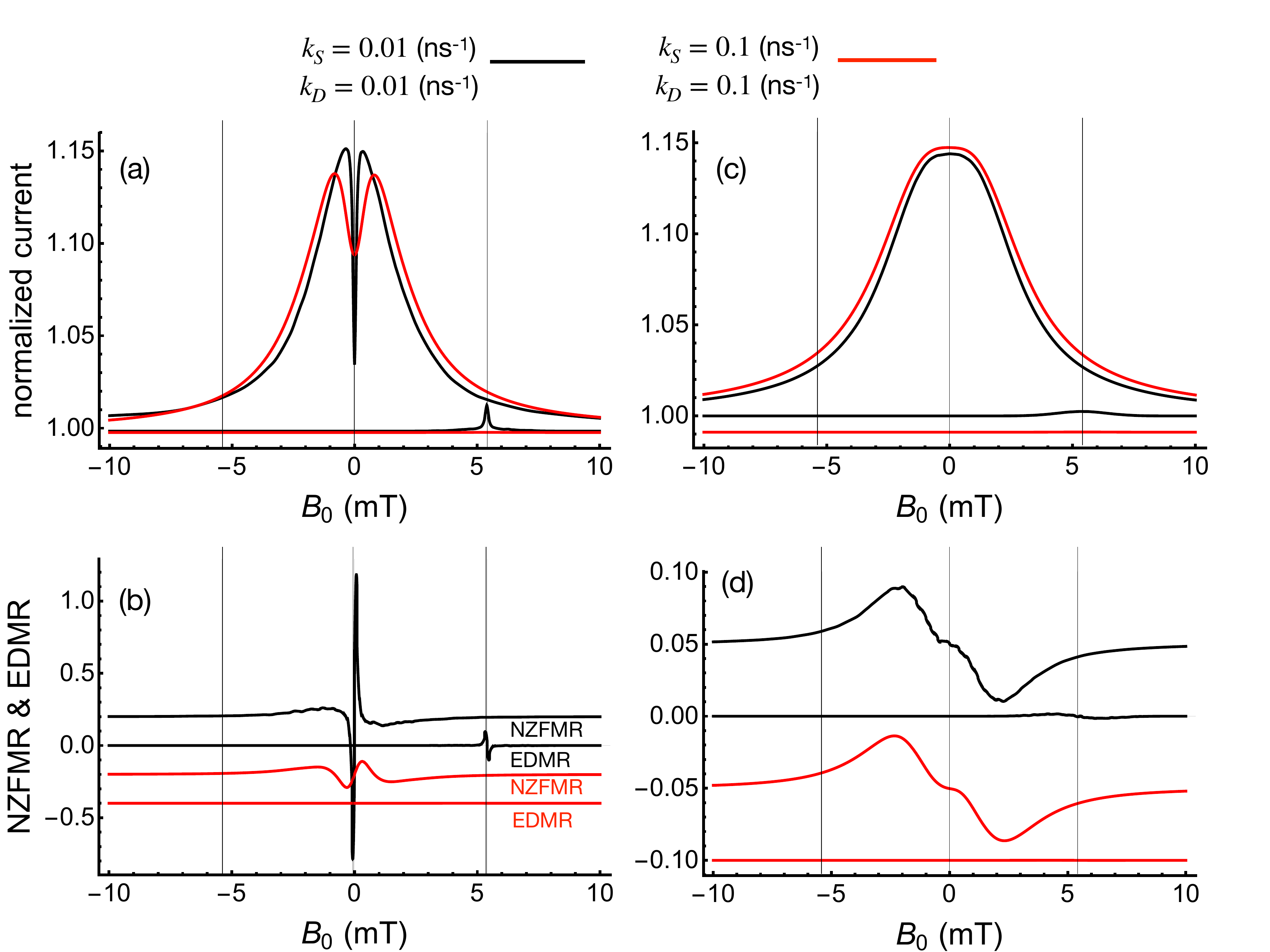}
        \caption[]
{Semiclassical approximation. (a,b) Two contrasting line shapes using  a single hyperfine interaction at $\bm{S}_1$ with $a_{Pb} = 1$ mT.
(c,d) Two contrasting line shapes using semiclassical nuclear spin at $\bm{S}_1$ and $\bm{S}_2$ using used $a_{Pb} = 1$ mT and $a_e = 1$ mT. For both calculations $g_e = 2$, $B_1 = 0.02$ mT, $ f= 150$ MHz ($\approx 5.4$ mT). The RWA and secular approximation (for the semiclassical term in the EDMR curve) are used in each calculation. The hyperfine constants here refer to the root-mean-square of the distribution of hyperfine fields; see Eq. (\ref{eq:gaussian}). (b) The red EDMR line is offset from the black EDMR line for the sake of visibility. (c,d) all four curves are offset from one another for visibility. To reduce computation time, EDMR is plotted for $B_0 > 0$; EDMR for $B_0 < 0$ shows identical behavior.}\label{fig:SClines} 
        \end{centering}
\end{figure*}

\subsection{Semiclassical Nuclear Spins}

Figure \ref{fig:SClines} graphs examples of NZFMR and EDMR when one or both spins interact with a large number of nuclear spins, justifying the semiclassical approximation for the hyperfine interaction.
Of particular interest is (c,d) where both spins are in the classical hyperfine field. The low-field effect is significantly reduced compared to (a,b). 
Figure \ref{fig:SClines} uses identical hyperfine couplings for the two electrons or holes but this is not expected to be physical since the two states are different. Again we find that the LFE is typically smaller in size (relative to the `wide shoulder' or WS) when hyperfine interactions exist at both sites.

To clarify what is meant by the semiclassical approximation -- not every defect spin will be in same nuclear environment since $^{29}$Si is only present for about 4.7 \% of the silicon atoms present. 4.7 \% of the defects will have a nuclear spin at their host site but a great many more will have nuclear spins in their vicinity and still interact with the nuclear spin. If the 12 nearest silicon atoms to the central defect are considered, there are 224 possible configurations (\emph{i.e.} ranging from zero $^{29}$Si to 13 $^{29}$Si); nearly half of those configurations possess at least one spinful nucleus.While the hyperfine interaction is strongest if the P$_b$ defect itself is $^{29}$Si, it is much more likely that a P$_b$ defect will experience a smaller hyperfine interaction from a next nearest neighboring $^{29}$Si nucleus and it is this higher likelihood that can dominate the response.\footnote{See Appendix for details on these probabilities.} Our semiclassical calculation ascribes an effective hyperfine constant to describe the entire ensemble of defects in their various $^{29}$Si (and hydrogen) arrangements.\footnote{We are making this argument based on the easier-to-visualize (111) surface; it is expected to hold true also for the (100) surface which is the surface in our experiments. }

\subsection{NZFMR and EDMR in Si/SiO$_2$ MOSFETs}

By comparing to measurements of spin-dependent recombination current at Si/SiO$_2$ interfaces in the remainder of this article, we find that the semi-classical approximation does a much better job of explaining the NZFMR and EDMR line shape features simultaneously. 

\section{Experiment}\label{sec:experiment}

Our experiments utilized a low field and frequency (5.4 mT and 151 MHz at resonance) custom-built spectrometer consisting of a Kepco BOP 50-2M bipolar power supply, a Lakeshore Cryogenics model 475 DSP Gaussmeter and temperature compensated Hall probe, a Agilent model 83732B synthesized signal generator, a Doty Scientific RF coil and 150 MHz tuning box, and a custom-built electromagnet with a separate pair of coils for magnetic field modulation. We utilize a Stanford Research Systems SR570 preamplifier for transimpedance amplification of the sample device current. We utilize LabVIEW software for magnetic field modulation, spectrometer control, and data acquisition and processing. Noise limits the feasibility of the measurement since the spin-dependent changes in device current are a very small fraction of the total current. Thus, the LabVIEW software also utilizes a virtual lock-in amplifier with frequency and phase sensitive detection of the magnetic field modulated device current. The resulting EDMR and NZFMR curves appear as approximate derivatives of the true spectra. All measurements were made at room temperature. 
The planar Si/SiO$_2$ MOSFET has a 7.5 nm thick SiO$_2$ gate dielectric with  $a \approx$ 41,000 $\mu$m$^2$ channel area and a shorted source and drain region.The gate of the total structure is comprised of 420 fingers each with a length of 1 $\mu$m and a width of 98 $\mu$m and a shorted source and drain region. 

\begin{figure}[ptbh]
 \begin{centering}
        \includegraphics[scale = 0.35,trim = 0 0 200 0, angle = -0,clip]{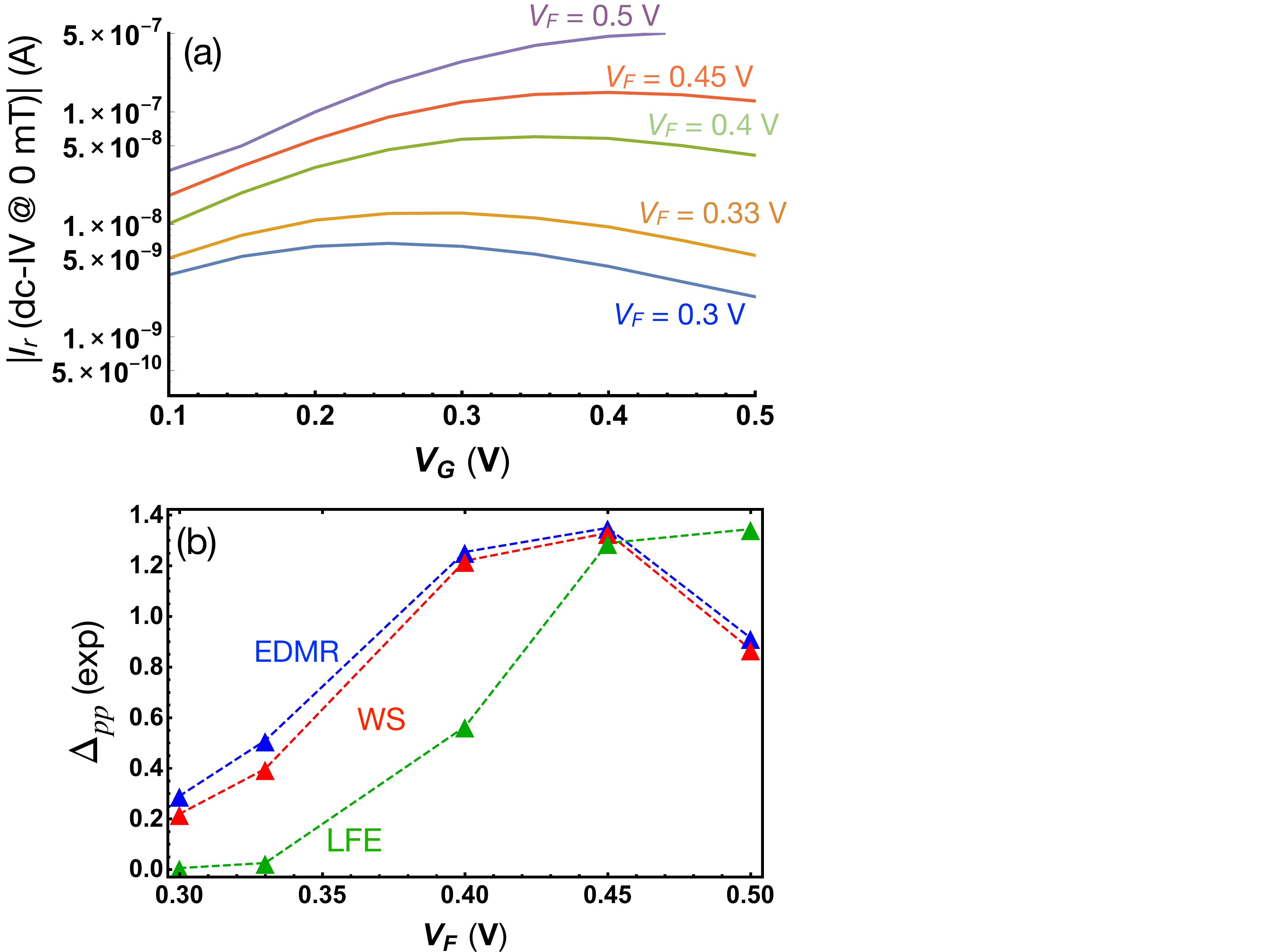}
        \caption[]
{(a) dc-IV curves as functions of gate voltage for several forward biases. The maximum recombination currents, at which our experiments are carried out, correspond to the maxima of each curve. (b) Peak-to-peak amplitudes of spin-dependent recombination (SDR) measurements.}  \label{fig:DCIV} 
        \end{centering}
\end{figure}

The oxidation was then followed up by a N$_2$ anneal at 925$^{\circ}$C. The devices were irradiated under a 0.2 V gate bias with a dose of 1 MRad. For our EDMR and NZFMR measurements, the source/drain to body junction biased at 0.3V, 0.33V, 0.4V, 0.45V, and 0.5V. with the gate bias held corresponding to the peak in the body current (Fig. \ref{fig:DCIV}(a)), with the body grounded; this biasing condition corresponds to the peak recombination current and optimal signal-to-noise. This biasing scheme is dc-IV and was first introduced by Grove.\cite{Grove1967} 
The recombination peak corresponds to the condition when electron and hole densities at the interface are equal. This condition is satisfied with the surface potential in depletion corresponding to $n = p = n_i \exp(q V_F/2kT) $. 

\begin{figure}[ptbh]
 \begin{centering}
        \includegraphics[scale = 0.3,trim = 0 7 1150 55, angle = -0,clip]{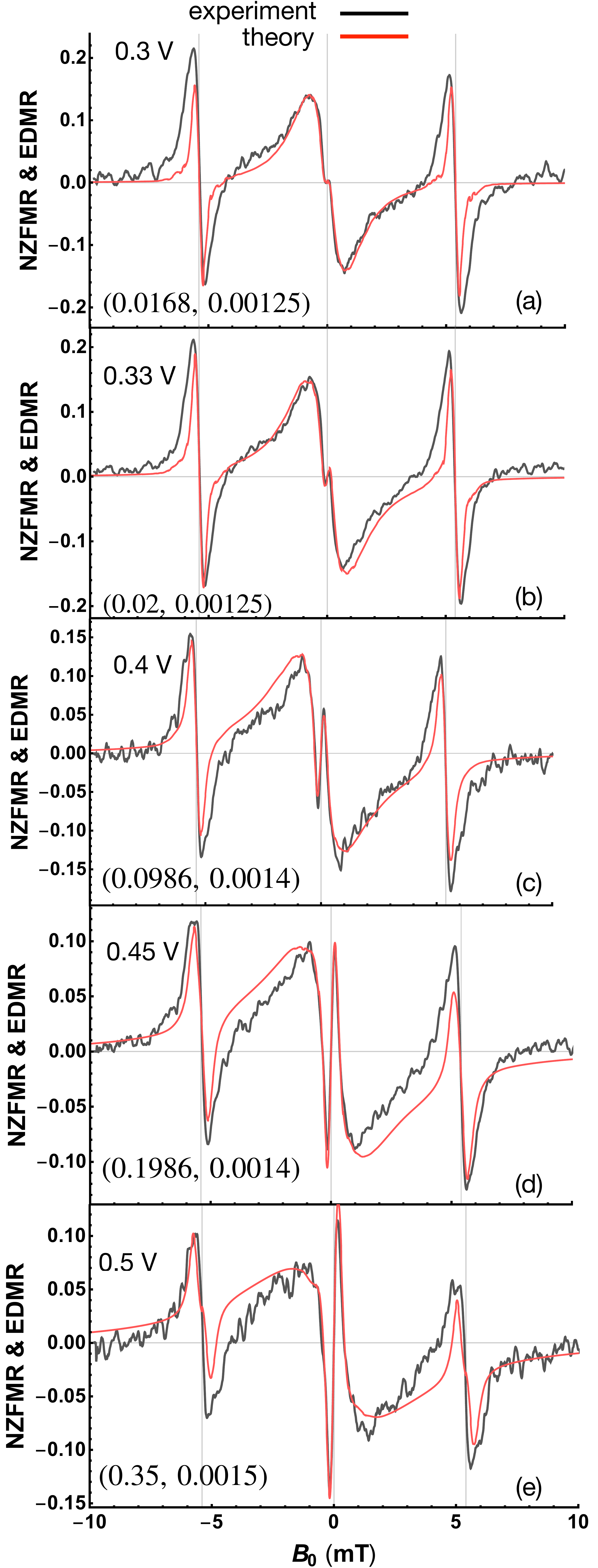}
        \caption[]
{Experimental and model NZFMR and EDMR line shapes. Experimental curves have been scaled by a constant factor to fit with model. Each panel displays the values of ($k_S$, $k_D$) used in the model in order to show qualitative comparisons. Hyperfine interactions are $a_e = 0.15$ mT and $a_{P_b} = 0.5$ mT. Other parameters are $g = 2$, $B_1 = 0.02$ mT, $ f= 150$ MHz ($\approx 5.4$ mT). The model curve in each panel averaged 20000 different nuclear field configurations.}  \label{fig:biasdependence} 
        \end{centering}
\end{figure}

The light red curves in Fig. \ref{fig:biasdependence} shows the post-irradiation spectrum. The pre-irradiation spectrum (not shown) exhibits neither an NZFMR or EDMR response. X-band measurements at ~9.4 GHz were also made on the same irradiated Si/SiO$_2$ MOSFETs, indicating that the near-interface traps are dominated by the P$_b$ center.\cite{Junipa1990}
Fig. \ref{fig:DCIV}(b) shows the peak-to-peak amplitudes of all three features seen in the experiment: EDMR occurring near $\pm 5.4$ mT, WS (wide shoulder), the broader shape (`shoulder') around zero field, and the LFE, the narrower feature (`dimple') around zero field. Note that, for the sake of comparing features to the theory, Fig. \ref{fig:biasdependence} does not display the actual values of the experimental amplitudes found in Fig. \ref{fig:DCIV}(b). 


\section{Discussion}

\begin{figure}[ptbh]
 \begin{centering}
        \includegraphics[scale = 0.45,trim = 0 220 560 3, angle = -0,clip]{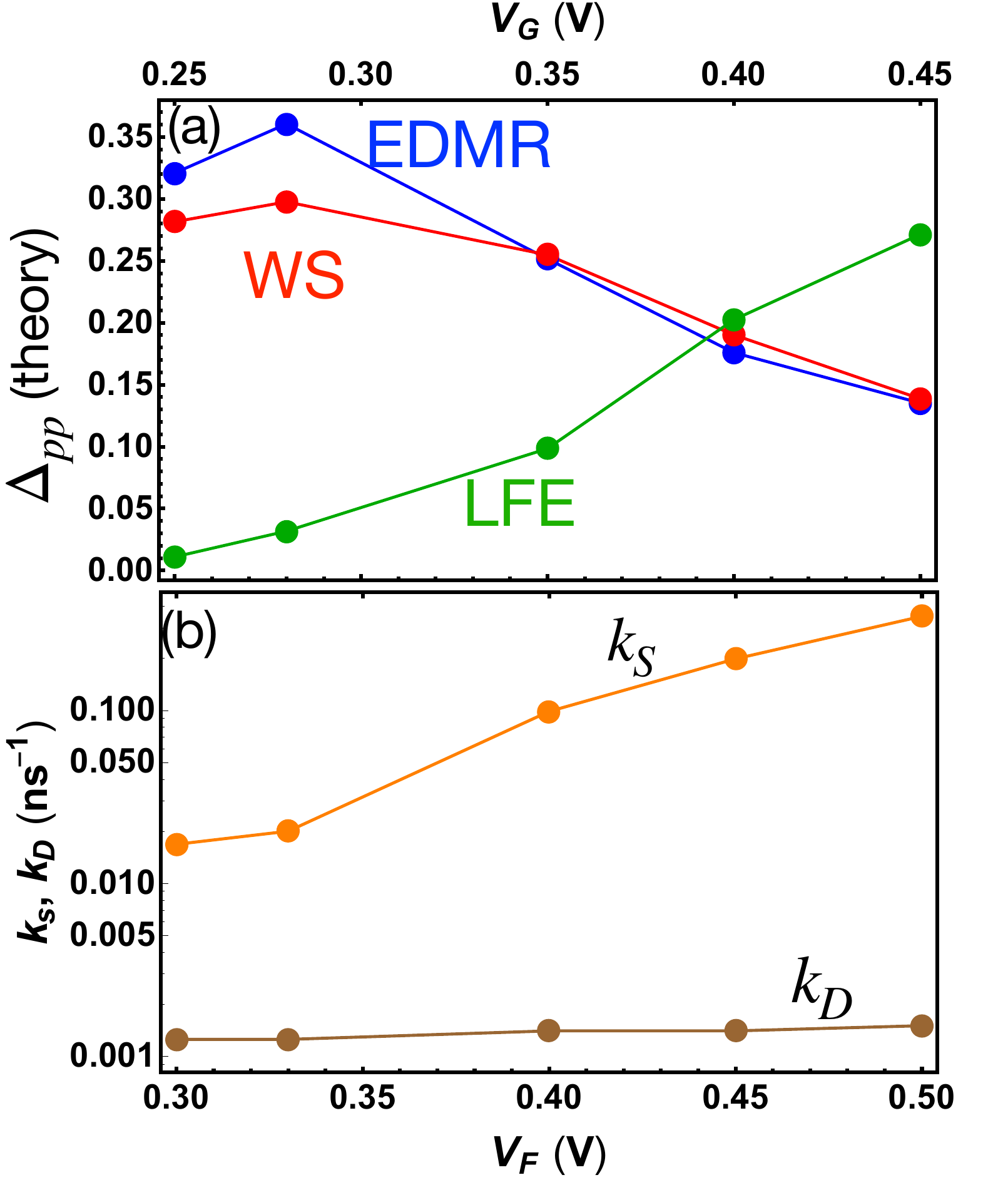}
        \caption[]
{(a) Calculation  peak-to-peak amplitudes in $d\rho_S/dB_0$ for EDMR, WS, and LFE features determined from Fig. \ref{fig:biasdependence}. (b) $k_S$ and $k_D$ model outcomes. }  \label{fig:thExp} 
        \end{centering}
\end{figure}

To offer an explanation of the experimental results of Fig. \ref{fig:biasdependence}, we first examine the lower forward biases of Fig. \ref{fig:biasdependence} (a) and (b). Fig. \ref{fig:biasdependence} (b) was considered in Ref. \onlinecite{Harmon2020}. We find that using the quantum model, neither  one nor two site hyperfine interactions can satisfactorily explain the NZFMR and EDMR simultaneously. For instance, an excellent fit to the NZFMR yields a minuscule EDMR peak. However by assuming semiclassical hyperfine interactions at both sites, NZFMR and EDMR relative features observed in our experiments can be satisfactorily recreated (red curves in Fig. \ref{fig:biasdependence} ). The absolute amplitudes of the experimental traces in Fig. \ref{fig:biasdependence} have been adjusted in order to compare the experimental and theoretical shapes. Figure \ref{fig:DCIV} (b) plots the actual experimental peak-to-peak measured values. The theory curves in Fig. \ref{fig:biasdependence}  are no longer normalized by the singlet probability at large field. Figure \ref{fig:thExp} (a) plots the EDMR, WS, and LFE peak-to-peak amplitude extracted from Fig. \ref{fig:biasdependence}. Figure \ref{fig:thExp} (b) displays the trends in $k_S$ and $k_D$ there were found when fitting the model to the measurements.

The semiclassical model for the defect spin is certainly sensible since each defect has a high likelihood of experiencing several Si hyperfine interactions. Moreover there is an abundance of nuclear spins originating from the passivating hydrogen at the surface for whose hyperfine couplings we do not possess detailed information.\cite{Himpsel1988, Brower1988, Tuttle1999} Given the range of hyperfine interactions from silicon (see Table I), the width of one half a milliTesla is not surprising.  The carrier spin also experiences a semiclassical interaction -- this is more apparent since the weakly bound electron will sample a larger number of nuclei within its localization radius. 
For the same reason, any hyperfine interaction felt would be expected to be less than that of the defect spin due to the less localized nature of the state. In light of the data that has been shown so far here, many hyperfine interactions at both spins, appears as the most likely possibility to explain both the NZFMR and EDMR relative line shape structure.

Recently, a paper by Frantz \emph{et al.} analyzed  spin dependent recombination NZFMR curves alone for the Si/SiO$_2$ MOSFET and spin-dependent trap assisted tunneling leakage current across the insulator in a hydrogenated amorphous silicon MIS capacitor.\cite{Frantz2020} By using the quantum model for the hyperfine interaction, accurate parameters for the hyperfine interactions and relative hyperfine abundances were extracted from a nonlinear least squares fitting routine. For the same set of parameters obtained from the fit for Si/SiO$_2$, the EDMR was negligible in size which is the same conclusion reached here.

This discrepancy can be explained by the following considerations that should be borne out by future in-depth analysis of the two approaches. While NZFMR and EDMR originate from same spin-dependent processes, their mechanics are different such that their dependencies on the various rates are not expected to be the same. A primary distinction is that EDMR requires an alternating field of magnitude $B_1$ to induce spin transitions while NZFMR does not. In the semiclassical model used here, the values for $k_S$ and $k_D$ were at least an order of magnitude smaller than the values found by Frantz \emph{et al.} using purely quantum nuclear spins when fitting the NZFMR line shape. As seen in Figs. \ref{fig:QMlines} and \ref{fig:SClines}, smaller $k_S$ and $k_D$ increase the relative size of EDMR in both the quantum and semiclassical calculations. The results here and those of Frantz \emph{et al.} suggest that the rates $k_S$ and $k_D$ are not constant but are drawn from a distribution of values.

\section{Analysis of Forward Bias Dependence}

\begin{figure}[ptbh]
 \begin{centering}
        \includegraphics[scale = 0.32,trim = 0 30 220 0, angle = -0,clip]{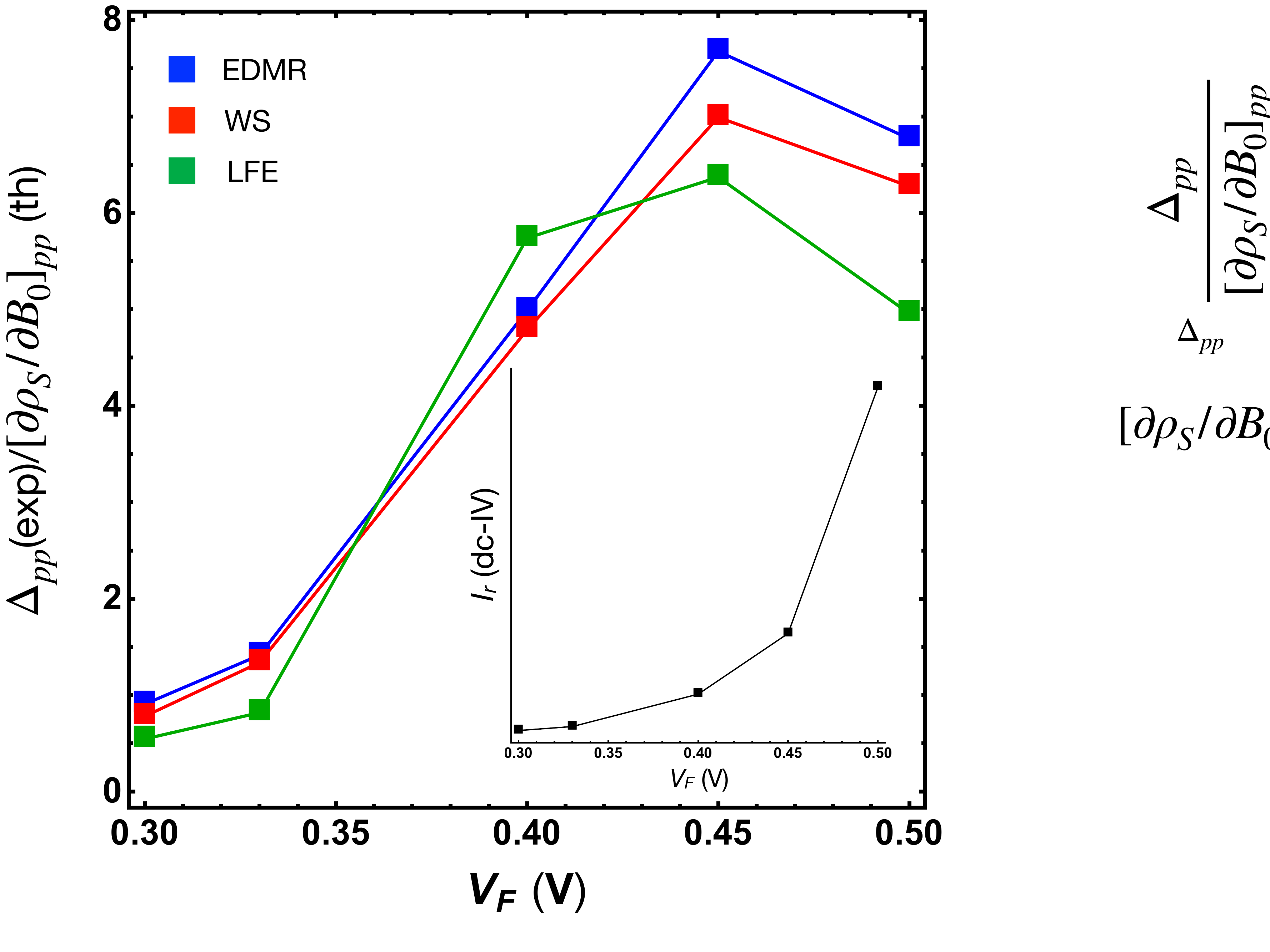}
        \caption[]
{Ratio of experimental peak-to-peak amplitudes to theoretical $[\partial \rho_S/\partial B_0]_{pp}$ for EDMR, WS, and LFE features. The inset shows the dc-IV recombination current measured against forward bias. We attribute increase in the ratio to the increase in dc-IV current for which the experiments take place.}  \label{fig:normalized} 
        \end{centering}
\end{figure}

The experimental line shape features are quantified by their peak-to-peak amplitudes, $\Delta_{pp}$. These amplitudes as a function of forward bias are shown in Figure \ref{fig:DCIV} (b). $\Delta_{pp}$ is related to the maximum recombination current by $\Delta_{pp} = \partial \rho_S/\partial B_0 * f(V_F)$ where $f(V_F)$ is an unknown function of forward bias. We expect $f(V_F)$ to scale in size with the dc-IV recombination current. This is tested in Fig. \ref{fig:normalized} by plotting the ratio 
\begin{equation}
\frac{\Delta_{pp}(\text{exp})}{[\partial \rho_S/\partial B_0]_{pp}(\text{th})}
\end{equation}
 which should yield $f(V_F)$. The main lines and inset line both increase with forward bias except at the largest forward bias; however in view of Fig. \ref{fig:DCIV}(b), there is uncertainty whether the maximum dc-IV recombination current was achieved at 0.5 V.
Without knowing the precise dependence of the field modulated device current on dc-IV, a more rigorous analysis of the relationship between the model and the experiment is not possible.

\section{Conclusion} 

The goal of this article has been to provide a sound theoretical basis for the near-zero-field magnetoresistance phenomena present at oxide-semiconductor interfaces in technologically relevant devices. The focus has been on spin-dependent recombination. We expect the theoretical approach to apply for spin-dependent trap-assisted transport (SDTAT) where NZFMR is also observed.\cite{Frantz2020}.
The model presented is able to qualitatively explain a series of spin-dependent recombination experiments where the gate and forward biases were changed.  To do so we demonstrated how a quantum model for the hyperfine interaction was insufficient to simultaneously describe the NZFMR and EDMR responses. Instead we utilized a semiclassical approximation for the hyperfine interaction which is reasonable considering the abundance of nuclear spins near the relevant spin pair.

\section{Acknowledgements}

The project or effort depicted was or is sponsored by the Department of the Defense, Defense Threat Reduction Agency under Grant HDTRA 1-18-1-0012 and Grant 1-16-0008. The content of the information does not necessarily reflect the position or the policy of the federal government, and no official endorsement should be inferred. 

\section{Data Availability}

The data that support the findings of this study are available from the corresponding author upon reasonable request.


%


\appendix

\begin{widetext}
\section{Derivation of the Rotating Wave Approximation Hamiltonian}

When transforming to a rotating coordinate system, either $\mathscr{H}_{a}$ or $\mathscr{H}_{b}$ become time independent while the other term obtains a relative $2\Omega$ precession rate that is routinely dropped within the so-called rotating wave approximation. 
\begin{equation}
\mathscr{H}_{1} \approx \mathscr{H}_{a} = B_1 ( g_e \mu_B S_{x,1} +  g_e \mu_B S_{x,2} -  g_n \mu_n \sum_j I_{x, j}) \cos\Omega t -  B_1  ( g_e \mu_B S_{y,1} +   g_e \mu_B S_{y,2} - g_n \mu_n \sum_j I_{y, j})\sin\Omega t,
\end{equation}
where the spin operators are dimensionless and $B_1 \ll B_0$. 

We can transform to a rotating coordinate system for which the transformed density matrix is $\tilde{\rho} = R^{-1} \rho R$ with $R = e^{i \Omega t (S_{z,1} + S_{z,2} + \sum_j I_{z, j})}$.
The motivation behind this transformation is to make the Hamiltonian time-independent.
The Zeeman Hamiltonians are
\begin{eqnarray}
\tilde{\mathscr{H}}_{0} = g_e \mu_B  B_0  (S_{z,1} + S_{z,2}) - g_n \mu_B  B_0  \sum_j I_{z, j},{}\\ \tilde{\mathscr{H}}_{1} = g_e \mu_B B_1 (S_{x,1} + S_{x,2}) - g_n \mu_B B_1 \sum_j I_{x, j}
\end{eqnarray}

Since $g_n \mu_n \ll g_e \mu_B$, the Zeeman Hamiltonians approximate as
\begin{equation}
\tilde{\mathscr{H}}_{0} = g_e \mu_B  B_0 (S_{z,1} + S_{z,2}) , ~~~~\tilde{\mathscr{H}}_{1} = g_e \mu_B B_1 (S_{x,1} + S_{x,2}).
\end{equation}

For the RWA to eliminate time dependence in the hyperfine term, some restrictions on the hyperfine terms must be assumed. All hyperfine coupling tensors must have the same principle axes \emph{and} possess axial symmetry (\emph{i.e.} $\hat{\bold{A}}_i = \text{diag}(a_{xx,i}, a_{xx,i}, a_{zz,i})$, $\hat{\bold{A}}_j = \text{diag}(a_{xx,j}, a_{xx,j}, a_{zz,j}$,  etc). 
When these conditions are satisfied,
\begin{equation}
    \tilde{\mathscr{H}}_{hf} = g_e \mu_B  \sum_j \bm{I}_j\cdot \hat{\bm{A}}_j \cdot \bm{S}_1
\end{equation}

The left-hand side of the Liouville equation becomes
\begin{equation}
\frac{\partial \tilde{\rho} }{\partial t} = \frac{\partial R^{-1} \rho R}{\partial t} = R^{-1} \rho\frac{\partial  R}{\partial t} + R^{-1}\frac{\partial \rho }{\partial t}R + \frac{\partial R^{-1} }{\partial t}\rho R = 
R^{-1} \rho R i \Omega (S_{z,1} + S_{z,2} + \sum_j I_{z, j})  + R^{-1}\frac{\partial  \rho }{\partial t}R - i \Omega (S_{z,1} + S_{z,2} + \sum_j I_{z, j}) R^{-1} \rho R.
\end{equation}
and then
\begin{equation}
\frac{\partial \tilde{\rho} }{\partial t} = \frac{\partial R^{-1} \rho R}{\partial t} = 
i \Omega [R^{-1} \rho R,  S_{z,1} + S_{z,2} + \sum_j I_{z, j} ] + R^{-1}\frac{\partial  \rho }{\partial t}R  = 
i \Omega [\tilde{\rho},  S_{z,1} + S_{z,2} + \sum_j I_{z, j} ] + R^{-1}\frac{\partial  \rho }{\partial t}R = 
-\frac{i}{\hbar}  [\hbar \Omega(S_{z,1} + S_{z,2} + \sum_j I_{z, j}), \tilde{\rho} ] + R^{-1}\frac{\partial  \rho }{\partial t}R.
\end{equation}
The last term we get from the transformed Liouville equation for which the coherent term can be determined from:
\begin{equation}
R^{-1}\frac{\partial  \rho }{\partial t}R = -\frac{i}{\hbar}R^{-1} [\mathscr{H},\rho]  R   = 
-\frac{i}{\hbar} (R^{-1}\mathscr{H} \rho R - R^{-1} \rho \mathscr{H}R) = 
-\frac{i}{\hbar} (R^{-1}\mathscr{H} R R^{-1} \rho R - R^{-1} \rho R^{-1} R \mathscr{H}R) = 
 -\frac{i}{\hbar} [\mathscr{\tilde{H}},\tilde{\rho}].  
\end{equation}
The remaining terms (stochastic ones) are determined similarly to yield a final expression:
\begin{equation}
\frac{\partial \tilde{\rho}}{\partial t} = - \frac{i}{\hbar}  [\hbar \Omega(S_{z,1} + S_{z,2} + \sum_j I_{z, j}), \tilde{\rho} ] -\frac{i}{\hbar} [\tilde{\mathscr{H}}_{0}+ \tilde{\mathscr{H}}_{hf} + \tilde{\mathscr{H}}_{1}  , \tilde{\rho}] - \frac{k_S}{2} \{ P_S, \tilde{\rho} \}  - \frac{k_T}{2} \{ P_T, \tilde{\rho} \}+ G \mathbb{1}.
\end{equation}
The two terms on the right hand side can be combined to make an effective Hamiltonian
\begin{equation}
\tilde{\mathscr{H}}_{} = g_e \mu_B  (B_0 + \frac{\hbar \Omega}{g_e \mu_B }) (S_{z,1} + S_{z,2}) - g_n \mu_B  (B_0 - \frac{\hbar \Omega}{g_n \mu_B }) \sum_j I_{z, j} +
g_e \mu_B  \sum_j \bm{I}_j\cdot \hat{\bold{A}}_j \cdot \bm{S}_1
 + g_e \mu_B B_1 (S_{x,1} + S_{x,2}) - g_n \mu_B B_1 \sum_j I_{x, j}
\end{equation}
\end{widetext}

\begin{figure}[ptbh]
 \begin{centering}
        \includegraphics[scale = 0.4,trim = 250 225 100 120, angle = -0,clip]{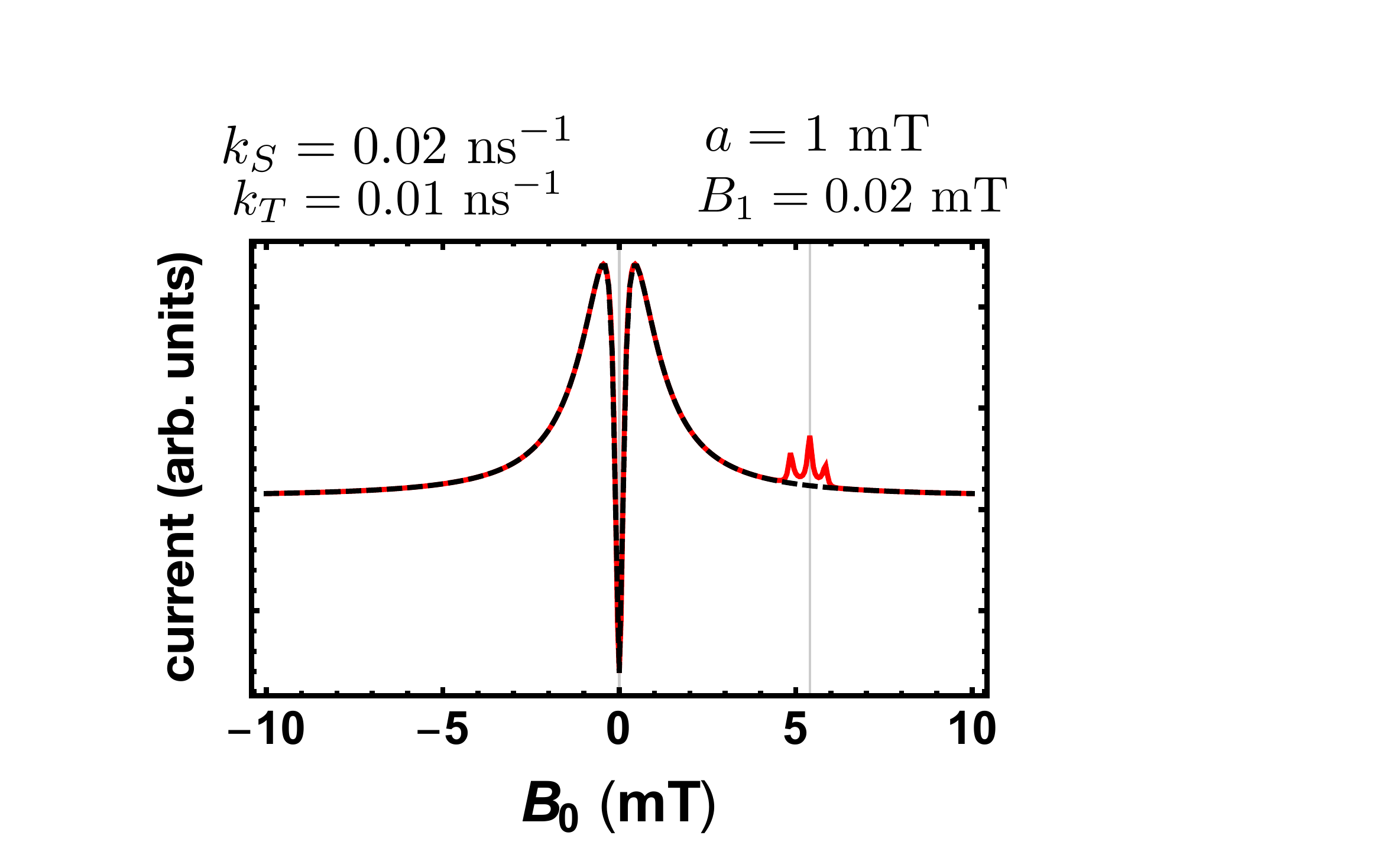}
        \caption[]
{The NZFMR line shape is calculated precisely whether the full EDMR or near-zero field Hamiltonians are used.}\label{fig:EDMRandNZFMRcomp} 
        \end{centering}
\end{figure}

For the range of parameters occurring in this article, using the RWA does not effect the NZFMR line shape as shown in Figure \ref{fig:EDMRandNZFMRcomp}.

\section{The P$_b$ defect at the S$\text{i}$/S$\text{i}$O$_2$ Interface}

\subsection{Defect-Hyperfine Statistics for S$\text{i}$/S$\text{i}$O$_2$ dangling bond defects}

The P$_b$ dangling bond defect at the interface of S$\text{i}$/S$\text{i}$O$_2$ is one of the better characterized defects. Experimental  and theoretical \cite{Edwards1987} treatments have determined its hyperfine structure so make this system an excellent testbed of our theory. The defect is shown in Figure \ref{fig:Pb0}.

The quickest approach is to assume that the defect atom dominates the magnetic field response of the recombination current. Only 4.7\% of silicons possess a spin-$\frac{1}{2}$ nucleus so to a first approximation only this subset is included. The hyperfine interaction is not isotropic as tabulated in Table \ref{tab:5/tc} where the first row is the hyperfine interaction with the defect center atom. For now neighboring silicons are neglected. 

Binomial probability distribution:
\begin{equation}
P(n, k, p) = {n \choose k}p^k(1 - p)^{n-k}
\end{equation}
which tells you the probability for $k$ of $n$ Si atoms to be the $^{29}$Si if the probability for any individual Si is $p\approx 0.047$.
So the probability of a specified defect being $^{29}$Si is $P_d(k_d) = P(1,k_d,p)$ which is 0.047 if $k_d = 1$.
But we are also interested whether nearby Si atoms have spinful nuclei since the defect wave function has some spatial overlap with, 
as we will see, its nearest and next nearest neighbors \cite{Edwards1987}. 
The hyperfine interactions are equivalent with its nearest neighbors when that nearest neighbor possesses nuclear spin.
Since there are three such atoms, the probability of $k_{nn}$ of those three being spinful is $P_{nn}(k_{nn}) = P(3,k_{nn},p)$.
As for next nearest neighbors (of which there are nine), there exist two different hyperfine couplings to the paramagnetic defect spin: three atoms are termed as bulk next nearest neighbors and the other  six  atoms are known as surface next nearest neighbors \cite{Edwards1987}.
The respective probabilities for spinful nuclei are  $P_{nnn, B}(k_{B}) = P(3,k_B,p)$ and  $P_{nnn, S}(k_B) = P(6,k_S,p)$.
The joint probability distribution for any number of the 13 relevant atoms to have spinful nuclei is
\begin{equation}
\mathcal{P}(k_d, k_{nn,} k_{S}, k_B) = P_d(k_d)  P_{nn}(k_{nn})  P_{nnn, B}(k_{B}) P_{nnn, S}(k_S),
\end{equation}
where $k_d \in \{0,1\}$, $k_{nn} \in \{0..3\}$, $k_B \in \{0..3\}$, and $k_S \in \{0..6\}$.
A given defect then is in one of 224 configurations --- $\sim$ 46\% of which possess at least one spinful nucleus at either the defect or its nearest/next nearest neighbors.

Furthermore, if hyperfine interactions exists for both electrons, the probability distribution must be expanded to include the probabilities of interaction at the second electron as well.

\subsection{Hyperfine interactions of P$_b$ defect}

Brower was the first to measure the P$_b$ hyperfine constants using EPR \cite{Brower1983}.
The measurement took place at 20 K and showed that the interaction was anisotropic.

\begin{equation}
\hat{\bold{A}} = a \mathbb{1} + \hat{\bold{T}},
\end{equation}
where
\begin{equation}
\hat{\bold{A}} = \left(
\begin{array}{ccc}
a_x & 0 &0\\
 0 & a_y & 0 \\
0 & 0 & a_z \\
\end{array}
\right)  ~~~ \text{and}~~~ \hat{\bold{T}}  = \left(
\begin{array}{ccc}
T_x & 0 &0\\
 0 & T_y & 0 \\
0 & 0 & T_z \\
\end{array}
\right) .
\end{equation}
All matrix elements are listed in Table \ref{tab:5/tc} as computed by Edwards in Ref. \onlinecite{Edwards1987}.
The experiment of Brower agree well with the ``Defect atom" numbers in the table.

\begin{widetext}
\begin{table*}[]
\addtolength{\tabcolsep}{3pt}    
\centering
\begin{tabular}{l c l  l l l l l l}
\hline
\hline
  & Number/defect  &  $a_x$ & $a_y$   & $a_z$  & $a$ & $T_x$ & $T_y$ & $T_z$\\
\hline
Defect atom & 1 &$ 9.65 $ & 9.65  & 15.72 & 11.67 & -2.023 & -2.023 & 4.046\\
Nearest neighbor (n.n) & 3 & $-0.399$ & $-0.093$  & $-0.072$ & 0.188 & -0.211 & 0.095 & 0.116\\
Bulk next n.n & 3 & $2.12$ & $2.18$ & $2.70$ & 2.33 & -0.214 & -0.153 & 0.367\\
Surface next n.n. & 6 & $-0.286$ & $-0.214$  & $-0.187$ & 0.229 & -0.057 & 0.015 & 0.042\\
\hline
\hline
\end{tabular}
\addtolength{\tabcolsep}{-3pt}
\caption{\label{tab:5/tc}Hyperfine matrix elements \cite{Edwards1987}. All values in mT. Note that some of these interactions do not possess axial symmetry in which case the RWA is invalid.}
\end{table*}
\end{widetext}

\newpage

\end{document}